\documentclass[
 reprint,
 amsmath,amssymb,
 aps,showkeys,superscriptaddress,pra
]{revtex4-2}

\usepackage{graphicx}
\usepackage{dcolumn}
\usepackage{bm}

\begin{document}


\title{Probing transverse-momentum- and pseudorapidity-dependent flow vector decorrelation in p--Pb collisions at the CERN Large Hadron Collider}



\author{Siyu Tang}
    \affiliation{School of Mathematical and Physical Sciences, Wuhan Textile University, Wuhan 430200, China}
    \affiliation{Shanghai Research Center for Theoretical Nuclear Physics, NSFC and Fudan University, Shanghai 200438, China}
\author{Zuman Zhang} 
    \affiliation{School of Physics and Mechanical Electrical \& Engineering, Hubei University of Education, Wuhan 430205, China}
    \affiliation{Institute of Theoretical Physics, Hubei University of Education, Wuhan 430205, China}
    \affiliation{Key Laboratory of Quark and Lapten Physics (MOE), Central China Normal University, Wuhan 430079, China}
\author{Chao Zhang}
    \email[Correspondence email address: ]{chaoz@whut.edu.cn}
    \affiliation{School of Physics and Mechanics, Wuhan University of Technology, Wuhan, 430200, China}
    \affiliation{Shanghai Research Center for Theoretical Nuclear Physics, NSFC and Fudan University, Shanghai 200438, China}
    \affiliation{Key Laboratory of Quark and Lapten Physics (MOE), Central China Normal University, Wuhan 430079, China}
\author{Liang Zheng}
    \email[Correspondence email address: ]{zhengliang@cug.edu.cn}
    \affiliation{School of Mathematics and Physics, China University of Geosciences (Wuhan), Wuhan 430074, China}
    \affiliation{Shanghai Research Center for Theoretical Nuclear Physics, NSFC and Fudan University, Shanghai 200438, China}
\author{RenZhuo Wan}
    \email[Correspondence email address: ]{wanrz@wtu.edu.cn}
    \affiliation{Hubei Key Laboratory of Digital Textile Equipment, Wuhan Textile University, Wuhan 430200, China}
    \affiliation{School of Electronic and Electrical Engineering, Wuhan Textile University, Wuhan 430200, China}  
\date{\today} 


\date{\today}

\begin{abstract}
    The event-by-event fluctuations in the initial energy density of the nuclear collisions lead to the decorrelation of second order flow vector, as well as its transverse-momentum ($p_{\mathrm{T}}$) and pseudorapidity ($\eta$) dependence as observed in high-energy heavy-ion collisions. Recent measurements at the CERN Large Hadron Collider shown that these decorrelations are also observed in small collision systems. In this work, a systematic study of the transverse-momentum- and pseudorapidity-dependent flow vector decorrelation is performed in p--Pb collisions at the 5.02 TeV with A Multi-Phase Transport (AMPT) model using different tunings of the initial conditions, partonic and hadronic interactions. It is found that the string-melting version of the AMPT model provides a reasonable description of the measured flow vector decorrelation as a function of $p_{\mathrm{T}}$ and $\eta$. We demonstrate that the hadronic scatterings do not have significant impact on decorrelation in p--Pb collisions for different centrality selections, while both initial conditions and partonic interactions influence the magnitude of the decorrelations. In addition, the nonflow, especially the long-range jet correlation, plays an important role in generating the decorrelation in small collision systems. The comparison of data and model presented in this paper provide further insights in understanding the fluctuations of the flow vector with $p_{\mathrm{T}}$ and $\eta$ in small collision systems and has referential value for future measurements.
\end{abstract}

\keywords{fluctuations, decorrelation, flow vector, small collision systems, LHC}

\maketitle

\section{Introduction}
High-energy heavy-ion collisions, such as those at the Relativistic Heavy-Ion Collider (RHIC) and the Large Hadron Collider (LHC)~\cite{Gyulassy:2004zy,STAR:2005gfr,PHENIX:2004vcz,Muller:2012zq}, offer an unique opportunity to study the nuclear matter under extreme conditions of temperature and density. These collisions create the quark-gluon plasma (QGP)~\cite{Shuryak:1978ij,Shuryak:1980tp}, a state of matter where quarks and gluons are no longer confined within hadrons but instead form a hot, dense medium. One of the most important observable for the formation of the QGP is the azimuthal anisotropy of produced particles, which is typically characterized by the Fourier expansion of the hadron yield distribution as a function of azimuthal angle $\varphi$:
\begin{equation}
\frac{dN}{d\varphi} \propto 1+2\sum_{n}^{\infty} v_{n}\mathrm{cos}[n(\varphi-\Psi_{n})],
\label{eq: flow defination}
\end{equation} 
where the Fourier coefficient $v_{n}$ and symmetry plane angle $\Psi_{n}$ represent the magnitude and orientation of the $n^{th}$ order flow vector $V_{n}\equiv v_{n}e^{in\Psi_{n}}$~\cite{Voloshin:1994mz,Poskanzer:1998yz}. Plenty of measurements focusing on the the second harmonic flow $v_{2}$ (elliptic flow), have been performed at the both RHIC~\cite{STAR:2000ekf,STAR:2013ayu,PHENIX:2003qra,STAR:2013qio,PHENIX:2015idk} and LHC~\cite{ALICE:2010suc,ALICE:2011ab,ALICE:2014wao,ALICE:2016ccg,ALICE:2017fcd,ATLAS:2012at,ATLAS:2011ah,ATLAS:2013xzf,CMS:2012xss,CMS:2012zex,CMS:2012tqw}. The comparisons to theoretical model calculations provided unprecedented constraints on the fundamental transport properties of the QGP medium~\cite{Heinz:2013th,Luzum:2013yya,Shuryak:2014zxa,Song:2017wtw}, such as the ratio of shear viscosity to entropy density ($\eta$/s), which indicate that the QGP created in heavy-ion collisions behaves as a nearly perfect fluid.

Due to the event-by-event fluctuations in the initial energy density of the nuclear collisions~\cite{Voloshin:2008dg,Heinz:2013th}, the fluctuations of second order flow vector $V_{2}$, as well as its transverse-momentum ($\it{p}_{\mathrm{T}}$) dependence have been shown in hydrodynamical models~\cite{Bozek:2018nne,Bozek:2021mov}. These fluctuations could result in the breakdown of factorization of two-particle angular correlations into a product of single-particle flow coefficients in different $\it{p}_{\mathrm{T}}$ intervals~\cite{Pang:2018zzo,Bozek:2018nne,Gardim:2017ruc,Zhao:2017yhj}, which were discovered in Pb--Pb collisions by the ALICE and CMS Collaborations~\cite{ALICE:2017lyf,CMS:2015xmx}. In addition, the flow vector fluctuations along the pseudorapidity ($\eta$) direction was investigated by hydrodynamic and parton transport models, where the factorization breakdown in $\eta$ was found to be sensitive to event-plane fluctuations at different $\eta$~\cite{Pang:2014pxa,Bozek:2017qir,Wu:2018cpc,Bozek:2015bna}. The CMS Collaboration measured the ratio of two-particle Fourier coefficients in rapidity bins $\eta$ and $-\eta$~\cite{CMS:2015xmx}, and the results show that the longitudinal fluctuations lead to a linear decrease of the ratio with $\eta$, i.e. the longitudinal flow decorrelations. These studies provide further constraint on the initial conditions and new insights to the longitudinal evolution of the medium formed in heavy ion collisions.

In recent years, the observation of non-zero $v_{2}$ in p--Pb and pp collisions raised the question of whether hydrodynamic flow also exist in these small collision systems~\cite{CMS:2012qk,ALICE:2012eyl,ATLAS:2012cix}. The extracted flow harmonics in p--Pb collisions have been studied in detail as a function of $\it{p}_{\mathrm{T}}$, $\eta$ and event multiplicity~\cite{Bzdak:2014dia,Tang:2023wcd,Tang:2024kot,Zhang:2022fum}. On the other hand, the fluctuations of flow vector as a function of $\it{p}_{\mathrm{T}}$ and $\eta$ are measured by the ALICE and CMS Collaborations~\cite{CMS:2015xmx,ALICE:2017lyf}. The effect of factorization breakdown was observed in p--Pb collisions and much smaller than that in Pb--Pb collisions. The measurements of longitudinal flow decorrelations were even extended to the smaller pp collisions by the ATLAS Collaboration~\cite{ATLAS:2023rbh}, and the results reveal significant sensitivity to the nonflow correlations. However, a systematic theoretical study for the $\it{p}_{\mathrm{T}}$- and $\eta$-dependent flow vector decorrelations in small collision systems is still missing. The effects from the initial and final state effects, as well as the nonflow contribution are also needed to be further investigated in the p--Pb collisions. Therefore, we present the first transport model study on this topic.

In this paper, the $\it{p}_\mathrm{T}$- and $\eta$-dependent flow vector decorrelation in p--Pb collisions are systematically studied with A Multi-Phase Transport (AMPT) model. The model with different tunings is applied to disentangle the impact of various physical processes, such as initial conditions, parton transport, nonflow contribution, and hadronic rescatterings. In section~\ref{The model}, a brief introduction about the AMPT and its various configurations is given. Section~\ref{Method} introduces the observables to characterize the flow vector decorrelation as a funcyion of the $\it{p}_{\mathrm{T}}$ and $\eta$, respectively. The two-particle correlation method and the advanced nonflow subtraction strategy are also presented. The results and related discussions are in section~\ref{Results}. Finally, a summary of this work is presented in section~\ref{summary}.  

\section{THE MODEL}
\label{The model}
The string-melting version of AMPT model~\cite{Lin:2004en,Lin:2021mdn} is employed in this work to study the flow vector decorrelation in p--Pb collisions at 5.02 TeV. The AMPT model consists four main stages: initial conditions, partonic scattering, hadronization, and hadronic rescattering. The heavy ion jet interaction generator (HIJING)~\cite{Gyulassy:1994ew} is incorporated in the model to generate the initial conditions, where minijet partons and soft-excited strings are produced and then converted to primordial hadrons based on Lund fragmentation. The strength of Lund fragmentation is controlled by two key parameter, Lund string parameters $a$ and $b$, which are approximately related to the string tension by $\kappa \propto \frac{1}{b(2+a)}$. With the string melting mechanism, primordial hadrons are converted into partons, which are determined by their flavor and spin structures. The subsequent parton interactions are treated with the Zhang’s parton cascade (ZPC) model~\cite{Zhang:1997ej}. Only elastic scattering between the partons is considered in the model, and the cross-section of two-body scattering is described by the following simplified equation:   
 \begin{equation}
\sigma = \frac{9\pi\alpha_{s}^{2}}{2\mu^{2}},
\label{eq: ZPC}
\end{equation} 
where $\alpha_{s}$ is the strong coupling and the $\mu$ is the Debye screening mass. Once the partons stop scattering, the nearest two or three quarks are combined into mesons or baryons using a quark coalescence model. The generated hadrons then enter the hadronic rescatterings process, which is described by an extended relativistic transport (ART) model~\cite{Li:1995pra} including both elastic and inelastic scatterings for baryon-baryon, baryon-meson, and meson-meson interactions. Finally hadronic scatterings are terminated at a cutoff time ($t_{\mathrm{max}}$), when the observables of interest are stable; a default cutoff time of $t_{\mathrm{max}}=30$ $\mathrm{fm}/c$ is used.

In order to separate the impact from different physical processes on the flow vector decorrelation, we varied the key parameters of the model in the calculations of the flow vector. The effect of the partonic phase is investigated by varying the partonic scattering cross section $\sigma$ from 0 mb and 0.5 mb to 3 mb. A smaller cross section corresponds to higher viscosity in viscous hydrodynamics in the AMPT model, and 0 mb of the cross section represents the exclusion of the parton scattering effect. Following the previous studies about the elliptic flow in p--Pb collisions, the default settings of Lund string parameters are $a=0.3$ and $b=0.15$. In this work, we further studied the initial conditions by varying $a=0.5$ and $b=0.9$, corresponding to a smaller string tension, when the partonic scattering cross section $\sigma$ is set to 0.5 mb. In addition, the cutoff time $t_{\mathrm{max}}$ is set to 0.6 fm/$\it{c}$ to turn off the hadronic rescatterings, while the resonance decays are still included. We also analyze the hadrons obtained right after the quark coalescence in the AMPT evolution ($t_{\mathrm{max}} = 0$ fm/$\it{c}$), where both hadronic rescatterings and resonance decays are disabled. All configurations are summarized in the table~\ref{tab: AMPT settings}. 
\begin{table}[htbp]
\centering
\begin{tabular}{|c|c|c|c|c|}
\hline
     & Lund a  & Lund b & Partonic cross section (mb) & $t_{\mathrm{max}}$ fm/$\it{c}$ \\ \hline
Par1 & 0.3     & 0.15   & 0.5                         & 30               \\ \hline
Par2 & 0.3     & 0.15   & 3                           & 30               \\ \hline
Par3 & 0.3     & 0.15   & 0                           & 30               \\ \hline
Par4 & 0.5     & 0.9    & 0.5                         & 30               \\ \hline
Par5 & 0.3     & 0.15   & 0.5                         & 0                \\ \hline
Par6 & 0.3     & 0.15   & 0.5                         & 0.6              \\ \hline

\end{tabular}
\caption{Table of different configurations of AMPT parameters used in this work.}
\label{tab: AMPT settings}
\end{table}

\section{OBSERVABLES AND METHOD}
\label{Method}
The flow vector decorrelation in p--Pb collisions is investigated by using the two-particle correlation method. To simplify, the azimuthal correlation between two emission particles can be represented by $N^{\mathrm{pair}}$ particle pairs as a function of the relative angle $\Delta\varphi = \varphi_{a} - \varphi_{b}$ between particles a and b and expanded in the Fourier series as follows:
\begin{small} \begin{equation}
\begin{aligned}
C(\Delta\varphi) = \frac{\mathrm{d}N^{\mathrm{pair}}}{d\Delta \varphi} \propto 1 + 2 \sum_{n=1}^{\infty}V_{n\Delta}(\it{p}_{\rm T}^{a},\it{p}_{\rm T}^{b})\mathrm{cos}[n(\Delta\varphi)],
\end{aligned}
\label{eq: Fourier 2PC} 
\end{equation} \end{small}
where $V_{n\Delta}$ refers to the two-particle $n$-th order harmonic. In a pure hydrodynamic scenario, the connection between the single- and two-particle harmonics can be expressed as:
\begin{small} \begin{equation}
\begin{aligned}
\langle V_{n\Delta}(p_{\mathrm{T}}^{a},p_{\mathrm{T}}^{b})\rangle = \langle v_{n}(p_{\mathrm{T}}^{a}) v_{n}(p_{\mathrm{T}}^{b}) \mathrm{cos}[n(\Psi_{n}(p_{\mathrm{T}}^{a})-\Psi_{n}({p_{\mathrm{T}}^{b}}))] \rangle ,
\end{aligned}
\label{eq: single phi distribution} 
\end{equation} \end{small}
where the bracket $\langle \cdot \rangle$ represents the average over all events, and the cosine term $\mathrm{cos}[n(\Psi_{n}(p_{\mathrm{T}}^{a})-\Psi_{n}({p_{\mathrm{T}}^{b}}))]$ shows the effects of the difference between the symmetry plane angle at $\it{p}_{\mathrm{T}}^{a}$ and $p_{\mathrm{T}}^{b}$ bin, due to the $\it{p}_{\mathrm{T}}$-dependent flow angle fluctuations. 

The tradition approach to calculate the $\it{p}_{\mathrm{T}}$-differential flow of particle $a$ (i.e. the particle of interest, POIs) is to firstly determine the flow coefficient of reference particles (RPs) over a wide kinematic range, called reference flow, and then the flow coefficient from $\it{p}_{\mathrm{T}}^{a}$ interval can be expressed as:
\begin{small} \begin{equation}
\begin{aligned}
v_{n}\{2\}(p_{\mathrm{T}}^{a}) &= \frac{V_{n\Delta}(p_{\mathrm{T}}^{a},\mathrm{ref})}{\sqrt{V_{n\Delta}(\mathrm{ref},\mathrm{ref})}} \\ 
  &= \frac{\langle v_{n}(p_{\mathrm{T}}^{a}) v_{n}^{\mathrm{ref}} \mathrm{cos}[n(\mathrm{\Psi}_{n}(p_{\mathrm{T}}^{a}) - \mathrm{\Psi}_{\mathrm{n}}  )] \rangle}{\sqrt{\langle {v_{n}^{\mathrm{ref}}}^2 \rangle}},
\end{aligned}
\label{eq: angle bracket} 
\end{equation} \end{small}    
where the $v_{n}(p_{\mathrm{T}}^{a}) v_{n}^{\mathrm{ref}}$ cannot be factorised into the product of $\sqrt{\langle v_{n}(p_{\mathrm{T}}^{a})^2 \rangle}$ and $\sqrt{\langle {v_{n}^{\mathrm{ref}}}^2 \rangle}$ if there are $p_{\mathrm{T}}$-dependent flow vector fluctuations. Another $\it{p}_{\mathrm{T}}$ differential flow observable, denoted as $v_{n}[2](p_{\mathrm{T}}^{2})$, was proposed in ref.~\cite{Heinz:2013bua}, which is not affected by the flow vector fluctuations:
\begin{small} \begin{equation}
\begin{aligned}
v_{n}[2](p_{\mathrm{T}}^{a})=\sqrt{V_{n\Delta}(p_\mathrm{T}^a, p_\mathrm{T}^a)} = \sqrt{\langle v_n(p_\mathrm{T}^a)^2\rangle}.
\end{aligned}
\label{eq: square bracket} 
\end{equation} \end{small} 
The ratio of $v_{n}\{2\}(p_{\mathrm{T}}^{a})$ and $v_{n}[2](p_{\mathrm{T}}^{a})$ is proposed to probe the $p_{\mathrm{T}}$-dependent flow vector decorrelation
\begin{small} \begin{equation}
\begin{aligned}
\frac{v_{n}\{2\}(p_{\mathrm{T}}^{a})}{v_{n}[2](p_{\mathrm{T}}^{a})}=\frac{\langle v_{n}(p_{\mathrm{T}}^{a}) v_{n}^{\mathrm{ref}} \mathrm{cos}[n(\mathrm{\Psi}_{n}(p_{\mathrm{T}}^{a}) - \mathrm{\Psi}_{\mathrm{n}}  )] \rangle}{\sqrt{\langle v_n(p_\mathrm{T}^a)^2\rangle} \sqrt{\langle {v_{n}^{\mathrm{ref}}}^2 \rangle}}.
\end{aligned}
\label{eq: v2 ratio} 
\end{equation} \end{small} 
One can see that if the $p_{\mathrm{T}}$-dependent flow vector fluctuations present, the ratio value is smaller than unity. Similarly, another observable to probe the $p_{\mathrm{T}}$-differential factorization ratio $r_n$ was proposed~\cite{Gardim:2012im,Heinz:2013bua},    
\begin{small} \begin{equation}
\begin{aligned}
r_n &= \frac{V_{n\Delta}(p_{\mathrm{T}}^{a},p_{\mathrm{T}}^{b})}{\sqrt{V_{n\Delta}(p_{\mathrm{T}}^{a},p_{\mathrm{T}}^{a})V_{n\Delta}(p_{\mathrm{T}}^{b},p_{\mathrm{T}}^{b})}} \\
    &= \frac{\langle v_{n}(p_{\mathrm{T}}^{a})v_{n}(p_{\mathrm{T}}^{b})\mathrm{cos}[n(\Psi_{n}(p_{\mathrm{T}}^{a})-\Psi_{n}(p_{\mathrm{T}}^{b}))] \rangle}{\sqrt{\langle v_{n}(p_{\mathrm{T}}^{a})^2 \rangle \langle v_{n}(p_{\mathrm{T}}^{b})^2 \rangle}}
\end{aligned}
\label{eq: rn} 
\end{equation} \end{small}
The $r_{n}$ < 1 indicates the breaking of factorization, suggesting the presence of $p_{\mathrm{T}}$-dependent flow vector fluctuations.

\begin{figure*}[!hbt]
\begin{center}
\includegraphics[width=2.\columnwidth]{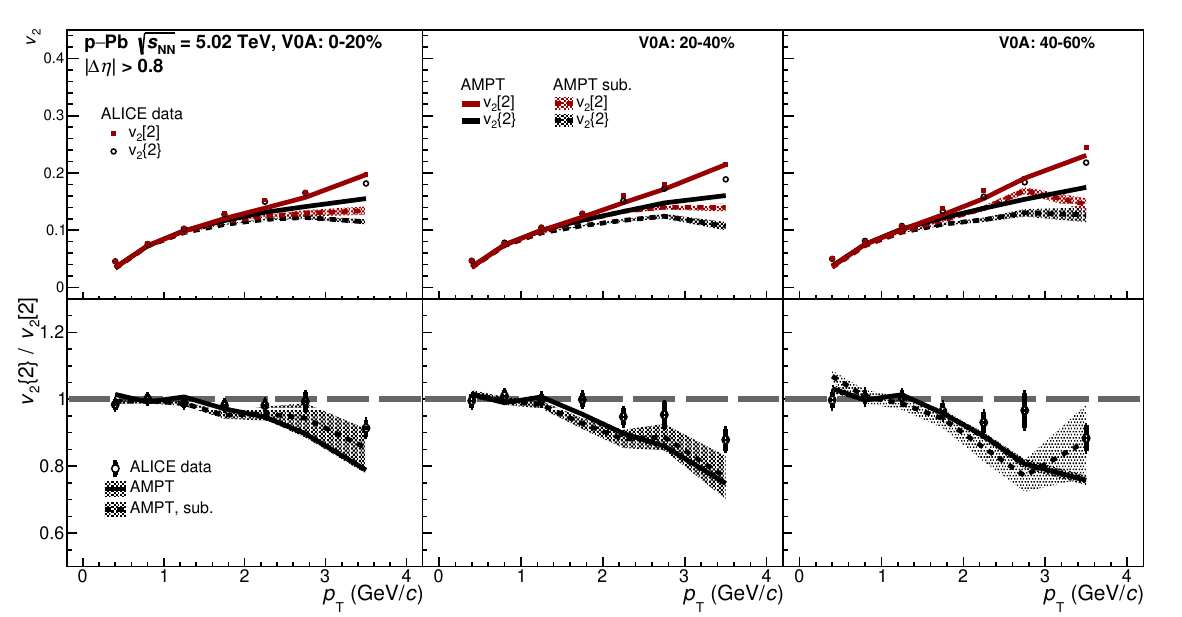}
\caption{(Color online) The $v_{2}\{2\}$ and $v_{2}[2]$ as a function of $p_{\mathrm{T}}$ and their ratio obtained from the AMPT model in 0--20\%, 20--40\% and 40--60\% p--Pb collisions at 5.02 TeV, in comparison of the ALICE measurements. The results with the nonflow subtraction obtained from the AMPT model are also shown.}
\label{Fig: v2 sub comp}
\end{center}
\end{figure*}

Similarly, the method to calculate the $\eta$-dependent flow is to construct the two-particle correlation between two different pseudo-rapidity bins, and the two-particle harmonics can be expressed as a formalism similar to Eq.~\ref{eq: single phi distribution} by replacing $p_{T}^{a}$ and $p_{T}^{b}$ by $\eta^{a}$ and $\eta^{b}$:
\begin{small} 
\begin{equation}
\begin{aligned}
\langle V_{n\Delta}(\eta^{a},\eta^{b}) \rangle = \langle v_{n}(\eta^{a})v_{n}(\eta^{b})\mathrm{cos}[n(\Psi_{n}(\eta^{a})-\Psi_{n}(\eta^{b}))] \rangle,
\end{aligned}
\label{eq: eta-factorization} 
\end{equation} 
\end{small}
where the $\mathrm{cos}[n(\Psi_{n}(\eta^{a})-\Psi_{n}(\eta^{b}))]$ represents the impact of the variation in symmetry plane angles between the $\eta^{a}$ and $\eta^{b}$ bins,  resulting from the $\eta$-dependent flow vector fluctuations. Then $\eta$-dependent factorization can be studied by the observable $r_n(\eta^{a},\eta^{b})$ proposed in Ref[], which is defined as
\begin{small} \begin{equation}
\begin{aligned}
r_n(\eta^{a},\eta^{b}) &= \frac{V_{n\Delta}(-\eta^{a},\eta^{b})}{V_{n\Delta}(\eta^{a},\eta^{b})} \\
                       &= \frac{v_{n}(-\eta^{a}) v_{n}(\eta^{b}) \langle \mathrm{cos}[n(\Psi_{n}(-\eta^{a})-\Psi_{n}(\eta^{b}))] \rangle}{v_{n}(\eta^{a}) v_{n}(\eta^{b}) \langle \mathrm{cos}[n(\Psi_{n}(\eta^{a})-\Psi_{n}(\eta^{b}))] \rangle }.
\end{aligned}
\label{eq: rn_eta} 
\end{equation} \end{small}
For symmetric collision systems like Pb--Pb collisions, the $v_{n}$ obtained from symmetric positive $[v_n(\eta_a)]$ and negative $[v_n(-\eta_a)]$ $\eta$ regions are identical after averaging over all events. Therefore, Eq.~\ref{eq: rn_eta} can be approximated by 
\begin{small} 
\begin{equation}
\begin{aligned}
r_n(\eta^{a},\eta^{b}) \approx \frac{\langle \mathrm{cos}[n(\Psi_{n}(-\eta^{a})-\Psi_{n}(\eta^{b}))] \rangle}{\langle \mathrm{cos}[n(\Psi_{n}(\eta^{a})-\Psi_{n}(\eta^{b}))] \rangle },
\end{aligned}
\label{eq: rn_eta_approx} 
\end{equation} 
\end{small}
where the $r_n(\eta^{a},\eta^{b})$ is less than unity if the factorization break down (i.e. the decorrelation presents). However, in an asymmetric collision system, such as p--Pb collisions, $v_n(\eta_a)$ and $v_n(-\eta_a)$ are generally not identical. As a result, it is not possible to isolate the $\eta$-dependent effects of symmetry-plane fluctuations in Eq.~\ref{eq: rn_eta}. The CMS collaboration proposed a method to take the product of $r_n(\eta^{a},\eta^{b})$ and $r_n(-\eta^{a},-\eta^{b})$ to remove the $v_{n}$ terms:
\begin{small}
\begin{equation}
\begin{aligned}
&\sqrt{r_n(\eta^{a},\eta^{b}), r_n(-\eta^{a},-\eta^{b})} \\
&\approx \sqrt{ \frac{\langle \mathrm{cos}[n(\Psi_{n}(-\eta^{a})-\Psi_{n}(\eta^{b}))] \rangle}{\langle \mathrm{cos}[n(\Psi_{n}(\eta^{a})-\Psi_{n}(\eta^{b}))] \rangle }  \frac{\langle \mathrm{cos}[n(\Psi_{n}(\eta^{a})-\Psi_{n}(-\eta^{b}))] \rangle}{\langle \mathrm{cos}[n(\Psi_{n}(-\eta^{a})-\Psi_{n}(-\eta^{b}))] \rangle }}.
\end{aligned}
\label{eq: rn_eta_approx_pPb} 
\end{equation} 
\end{small}
In this case, the $\eta$-dependent flow vector decorrelation in p--Pb collisions can also be studied.

\begin{figure*}[!hbt]
\begin{center}
\includegraphics[width=2.\columnwidth]{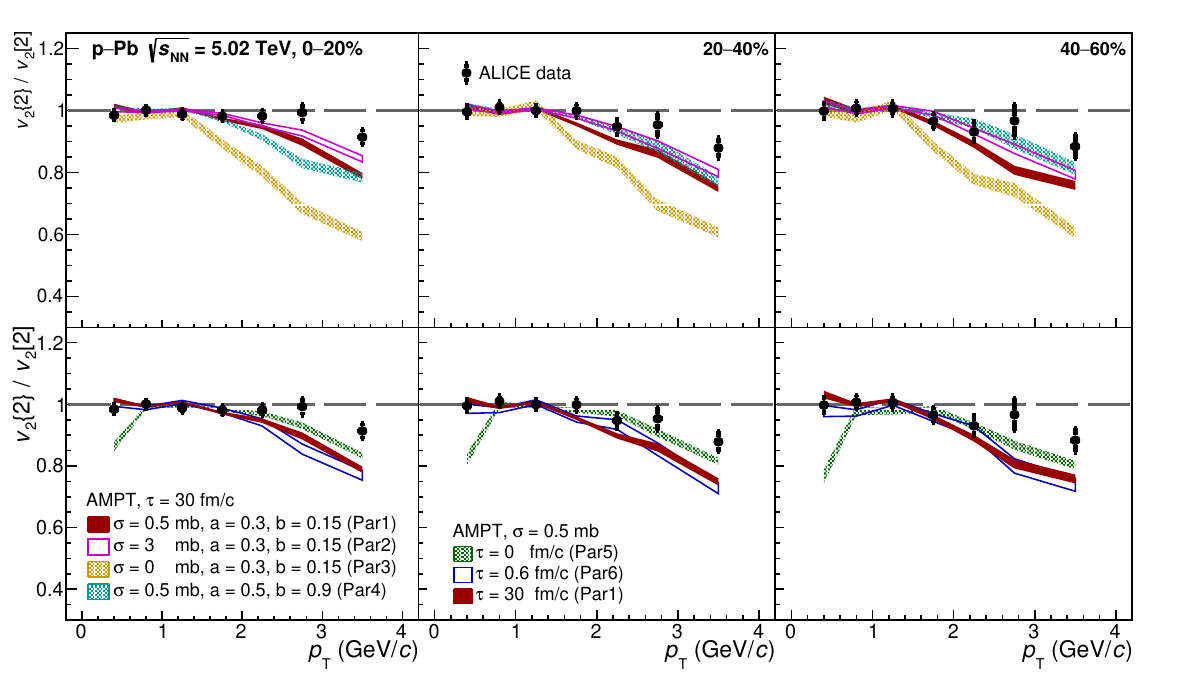}
\caption{(Color online) The $p_{\mathrm{T}}$-dependent $v_{2}\{2\}$/$v_{2}[2]$ ratio in 0--20\%, 20--40\%, 40--60\% centrality classes from the AMPT model with different configurations. ALICE data points are shown as black open circles for comparison. }
\label{Fig: Pt Decorr}
\end{center}
\end{figure*}

\begin{figure*}[!hbt]
\begin{center}
\includegraphics[width=1.48\columnwidth]{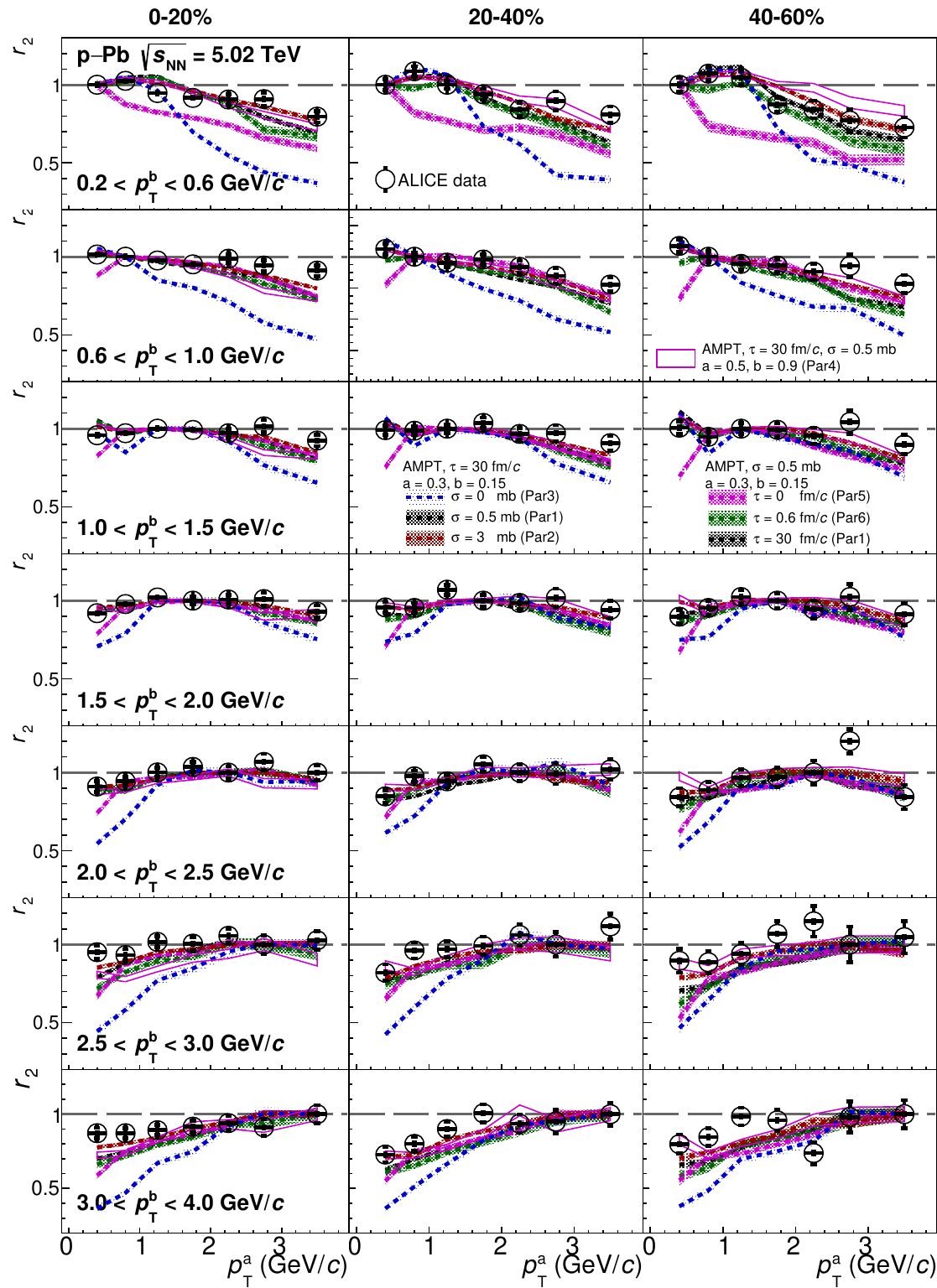}
\caption{(Color online) The factorization ratio $r_2$ as a function of $p^{a}_{\mathrm{T}}$, in different centrality and $p^{b}_{\mathrm{T}}$ ranges, obtained from the AMPT model with various configurations. ALICE data points are shown as black open circles for comparison.}
\label{Fig: r2 Decorr}
\end{center}
\end{figure*}

In small collision systems, the main contributions of the nonflow is from the jets, including both the short-range and long-range (also called "recoil") jet correlations. The former can be effectively removed by introducing a large rapidity gap between the trigger and associated particles during the construction of the correlations, and the latter can be suppressed by many methods in the measurements based on the different assumptions. In this work, the template fit method developed by the ATLAS collaboration is applied, which has been proven effective for nonflow subtraction in our previous studies in the AMPT. For the given $\it{p}_{\mathrm{T}}$ and $\eta$ integral, the correlation function distribution obtained in high-multiplicity events is assumed to result from the superposition of the distribution obtained in low-multiplicity events scaled up by a multiplicative factor $F$ and a constant modulated by $\mathrm{cos}(n\Delta\varphi)$ for $n > 1$, as shown in
\begin{equation}
\begin{aligned}
C(\Delta\varphi) = FC^{\mathrm{LM}}(\Delta\varphi) + G(1 + 2 \sum_{n=1}^{3}V_{n\Delta}\mathrm{cos}(n\Delta\varphi)),
\end{aligned}
\label{eq: template fit}
\end{equation}
where $G$ denotes the normalization factor, and $V_{n\Delta}$ is the two-particle $n$-th order harmonic after the nonflow subtraction. The $C^{\mathrm{LM}}$ is the correlation distributions obtained from the low-multiplicity events, and in this work the collisions with 60--100\% centrality are selected. According to replacing Eq.~\ref{eq: Fourier 2PC} with Eq.~\ref{eq: template fit}, the flow vector fluctuations after the long-range nonflow subtraction can be calculated. 

\section{RESULTS AND DISCUSSION}
\label{Results}
\subsection{Transverse-momentum dependence of decorrelation}
We first examine the $p_{\mathrm{T}}$-differential $v_{2}$ of charged hadrons in the pseudorapidity region $|\eta|<$ 0.8 using the AMPT model with the default configuration (Par1) as described in Table~\ref{tab: AMPT settings}. The $v_{2}\{2\}$ and $v_{2}[2]$ are calculated according to Eq.~\ref{eq: angle bracket} and Eq.~\ref{eq: square bracket}. Figure~\ref{Fig: v2 sub comp} shows the $p_{\mathrm{T}}$-differential $v_{2}\{2\}$ and $v_{2}[2\}$ for 0--20\%, 20--40\%, and 40--60\% centrality bins obtained from the AMPT calculations, compared with ALICE measurements. To suppress contributions from short-range jet correlations, we apply $|\Delta\eta| > 0.8$. The calculations reasonably reproduce the $v_{2}$ data from central to semi-central collisions, and the ratio $v_{2}\{2\}/v_{2}[2\}$ shows deviations from unity above $\it{p}_{\mathrm{T}} \simeq$ 2 GeV/$c$. These deviations increase with $\it{p}_{\mathrm{T}}$, similar to findings in Pb--Pb collisions, indicating $p_{\mathrm{T}}$-dependent flow vector fluctuations in p--Pb collisions from the AMPT model. Unlike Pb--Pb collisions, these deviations do not significantly depend on centrality selections. To determine if these deviations are caused by non-flow effects from long-range jet correlations, we apply the template fit subtraction method described in Eq.~\ref{eq: template fit}. As expected, the $v_{2}$ after subtraction, shown in Fig.~\ref{Fig: v2 sub comp} (dashed line), is lower than the results without subtraction, especially at high $p_{\mathrm{T}} >$ 2 GeV/$c$. The ratio $v_{2}\{2\}/v_{2}[2\}$ after subtraction is consistent with that before subtraction within uncertainties. This indicates that the long-range jet contribution to the decorrelations is negligible here, and the observed $p_{\mathrm{T}}$-dependent flow vector fluctuations in p--Pb collisions are mainly from real flow signals.

\begin{figure*}[!hbt]
\begin{center}
\includegraphics[width=2.\columnwidth]{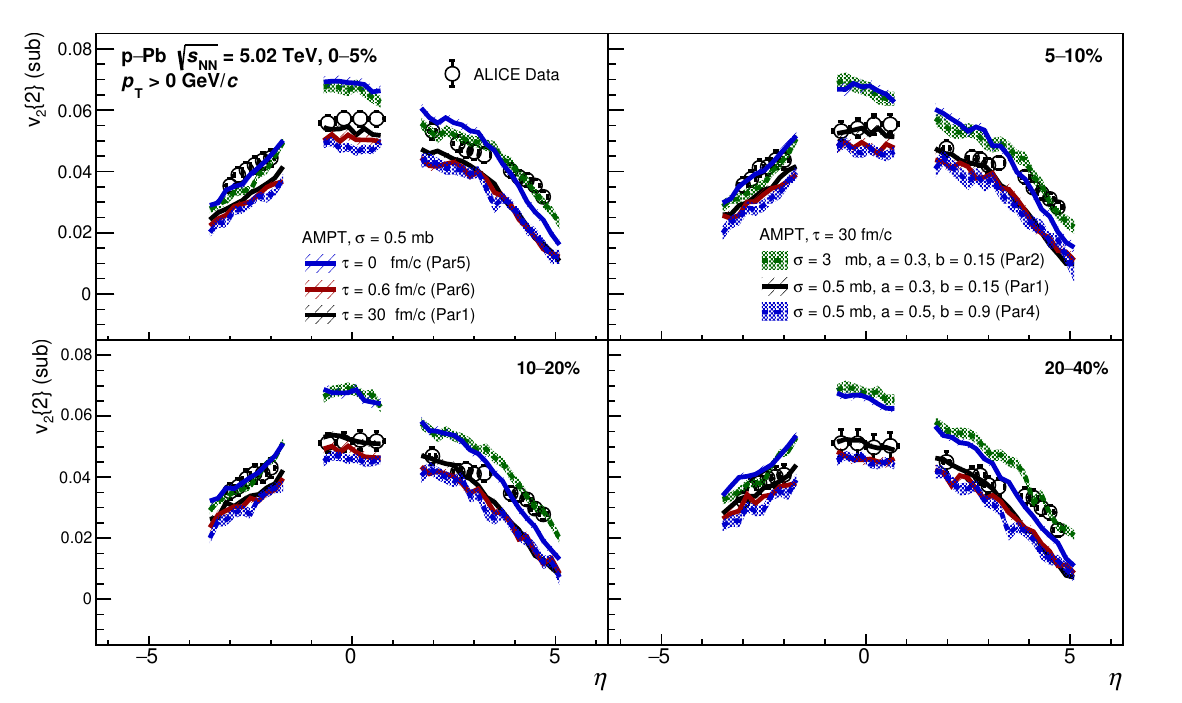}
\caption{(Color online) The $v_{2}$ as a function of pseudorapidity $\eta$ in in different centrality ranges, obtained from the AMPT model with various configurations. ALICE data points are shown as black open circles for comparison.}
\label{Fig: v2 eta}
\end{center}
\end{figure*}

The effect from varying the initial conditions via the Lund string parameters, $a$ and $b$, can be seen by comparing the results from Par4 and Par1, as shown in Fig.~\ref{Fig: Pt Decorr} (top). One can see that the $v_{2}\{2\}$/$v_{2}[2]$ ratio obtained with larger Lund $a$ and $b$ (Par4) is lower in the most 0--20\% collisions. With the increainsg of the centralities, the resuts from Par4 are enhanced, and finally are hihger than the results from Par1. The effects from the hadronic scatterings are studied by varying the cutoff time of hadronic rescatterings, which can seen in the comparison of the results of Par1, Par5 and Par6, as shown in Fig.~\ref{Fig: Pt Decorr} (bottom). The $v_{2}\{2\}$/$v_{2}[2]$ ratio from Par1 and Par6 are consistent within uncertainties from the central to peripheral collisions, indicating that the impact from hadronic scatterings on the flow vector fluctuations are negligible. It is consistent with the previous findings that the hadronic scatterings have almost no effects on the $p_{\mathrm{T}}$-differential $v_{2}$ in p--Pb collisions~\cite{Tang:2023wcd}. For the case of Par5, where both the resonance decay and the hadronic scatterings are turn off, the $v_{2}\{2\}$/$v_{2}[2]$ ratio is lower at $p_{\mathrm{T}}<0.8$ GeV/$c$ but higher at $p_{\mathrm{T}}>1.4$ GeV/$c$ compare to the results of Par1 and Par6. It is expected since the decay processes convert high-$p_{\mathrm{T}}$ hadrons into low-$p_{\mathrm{T}}$ particles. One can see that the results of all configurations underestimate the data, suggesting that the model needs further tuning, particularly of the initial conditions and parton scattering processes. 
The effect from varying the initial conditions via the Lund string parameters, $a$ and $b$, can be seen by comparing the results from Par4 and Par1, as shown in Fig.~\ref{Fig: Pt Decorr} (top). One can see that the $v_{2}\{2\}$/$v_{2}[2]$ ratio obtained with larger Lund $a$ and $b$ (Par4) is lower in the most 0--20\% collisions. With the increainsg of the centralities, the resuts from Par4 are enhanced, and finally are hihger than the results from Par1.
The effects from the hadronic scatterings are studied by varying the cutoff time of hadronic rescatterings, which can seen in the comparison of the results of Par1, Par5 and Par6, as shown in Fig.~\ref{Fig: Pt Decorr} (bottom). The $v_{2}\{2\}$/$v_{2}[2]$ ratio from Par1 and Par6 are consistent within uncertainties from the central to peripheral collisions, indicating that the impact from hadronic scatterings on the flow vector fluctuations are negligible. It is consistent with the previous findings that the hadronic scatterings have almost no effects on the $p_{\mathrm{T}}$-differential $v_{2}$ in p--Pb collisions. For the case of Par5, where both the resonance decay and the hadronic scatterings are turn off, the $v_{2}\{2\}$/$v_{2}[2]$ ratio is lower at $p_{\mathrm{T}}<0.8$ GeV/$c$ but higher at $p_{\mathrm{T}}>1.4$ GeV/$c$ compare to the results of Par1 and Par6. It is expected since the decay processes convert high-$p_{\mathrm{T}}$ hadrons into low-$p_{\mathrm{T}}$ particles. One can see that the results of all configurations underestimate the data, suggesting that the model needs further tuning, particularly of the initial conditions and parton scattering processes.         

The factorization ratio $r_{2}$, defined in Eq.~\ref{eq: rn}, as a function of $p^{a}_{\mathrm{T}}$ in different centrality and $p^{b}_{\mathrm{T}}$ ranges, is shown in Fig.~\ref{Fig: r2 Decorr}. The comparison between the model calculations from different AMPT configurations and the ALICE data is also shown. The model with a 0.5 mb cross section, $a$ = 0.3, $b$ = 0.15 and $\tau$ = 30 fm/$c$ (Par1), provides a fair description of the data in all centrality and $p_{\mathrm{T}}$ bins. The deviation of the $r_{2}$ from the unity has no dependence on the centrality selections, and is more pronounced with the increasing $|p^{a}_{\mathrm{T}}-p^{b}_{\mathrm{T}}|$. The $r_{2}$ obtained with larger Lund $a$ and $b$ (Par4) are higher in more peripheral collisions, reflecting the effects from the initial conditions. Changing the partonic scattering cross section from 0.5 mb to 3 mb (Par1 vs Par3) slightly enhance the $r_{2}$, while the results obtained with $\sigma=0$ mb shows a significant breakdown of the factorization at large $|p^{a}_{\mathrm{T}}-p^{b}_{\mathrm{T}}|$. 
The impacts from the hadronic scatterings are negligible for all centralities and $p_{\mathrm{T}}$ bins (Par1 vs Par6), but the resonance decay process still has large impacts on the $r_{2}$ at low-$p_{\mathrm{T}}$ region (Par1 vs Par5). One can see that the calculations of $r_{2}$ and $v_{2}\{2\}$/$v_{2}[2]$ ratio have generally similar dependence on the initial conditions, parton scattering cross section, hadron scattering process. It is expected since $r_{2}$ is basically the double-differential $v_{2}\{2\}$/$v_{2}[2]$ ratio.  

\begin{figure*}[!hbt]
\begin{center}
\includegraphics[width=2.\columnwidth]{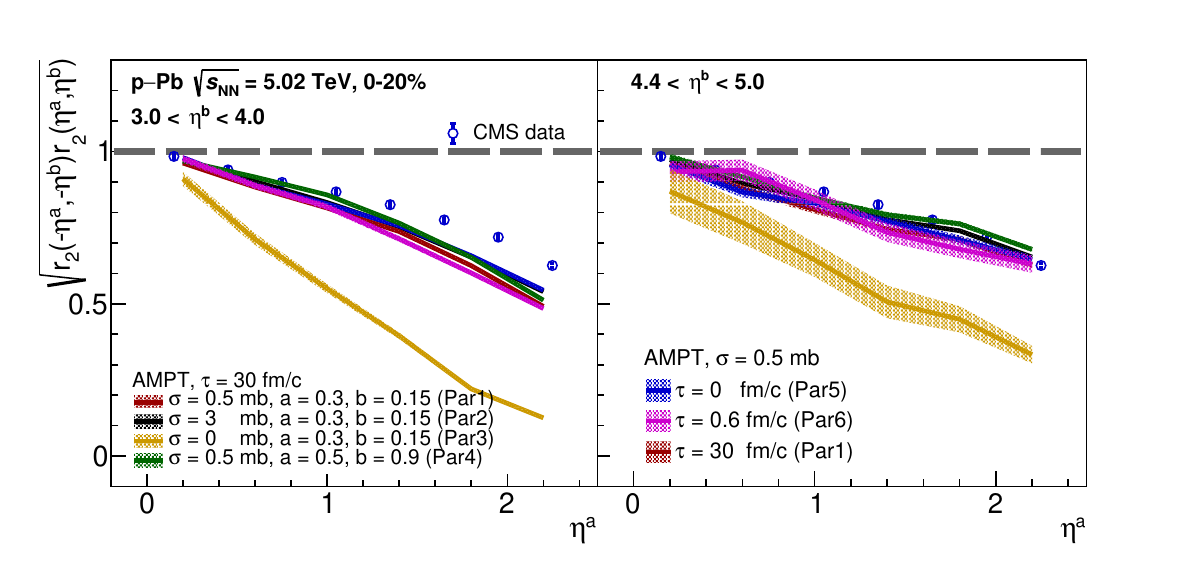}
\caption{(Color online) The square root of the product of factorization ratios, $\sqrt{r_{2}(\eta^{a},\eta^{b})r_{2}(-\eta^{a},-\eta^{b})}$, as a function of $\eta^{a}$ for 3.0 $< \eta^{b} <$ 4.0 (left) and $4.4 < \eta^{b} < 5.0$ (right), in 0--20\% centrality class of p--Pb collisions at 5.02 TeV, obtained from the AMPT model with various tunings. The CMS data points are shown as black open circles for comparison.}
\label{Fig: eta decorr}
\end{center}
\end{figure*}

\subsection{Pseudorapidity-dependence of decorrelation}
The $\eta$-dependent $v_{2}$ is first calculated before studying the decorrelation. We follow the two-particle correlation method used by the ALICE experiment. The correlation between three groups of particles is constructed based on Eq.~\ref{eq: template fit}, then the flow coefficients are extracted for three combinations:
\begin{equation}
\begin{aligned}
v_{n}(\eta^{a}) = \sqrt{\frac{V_{n\Delta}(\eta^{a},\eta^{b})V_{n\Delta}(\eta^{a},\eta^{c})}{V_{n\Delta}(\eta^{b},\eta^{c})}},
\end{aligned}
\label{eq: 3x2PC} 
\end{equation}
where $a$, $b$, and $c$ represent the hadrons with $p_{\mathrm{T}}>$ 0 GeV/$c$ at different rapidity bins. Similarly, $v_{n}(\eta^{a})$ and $v_{n}(\eta^{b})$ can be obtained by alternating the indices $a$, $b$, and $c$ in Eq.~\ref{eq: 3x2PC}. Figure~\ref{Fig: v2 eta} shows the $v_{2}$ of the charged hadrons for $p_{\mathrm{T}}>$ 0 GeV/$c$ as a function of $\eta$ in 0--5\%, 5--10\%, 10--20\%, and 20--40\% centrality ranges. The results obtained from different configurations of the AMPT model are compared with the ALICE data. The calculations obtained with a 0.5 mb cross section, $a$ = 0.3, $b$ = 0.15, and $\tau$ = 30 fm/$c$ (Par1) provide a good description of the data at mid-rapidity but slightly underestimate the $v_{2}$ at forward and backward rapidity. On the other hand, the Par2 calculations with a larger cross section ($\sigma$ = 3 mb) describe the data at forward and backward rapidity but significantly overestimate the $v_{2}$ at mid-rapidity. The calculations without hadron rescatterings (Par6) are lower compared to the Par1 results, especially at mid-rapidity, while the deviation is negligible at forward and backward rapidity. Similar behavior is observed in the results obtained with larger Lund $a$ and $b$ (Par4). Furthermore, the results obtained from Par5 are significantly enhanced for all rapidity bins, indicating the impact of resonance decay on the integrated-$v_{2}$. One can see that the results from all configurations cannot simultaneously describe the data at mid- and forward/backward-rapidity. This suggests that the rapidity dependence of parameters (e.g., the initial conditions, the strength of parton interactions) needs to be considered in further developments of the model~\cite{Lin:2021mdn}.   

\begin{figure*}[!hbt]
\begin{center}
\includegraphics[width=2.\columnwidth]{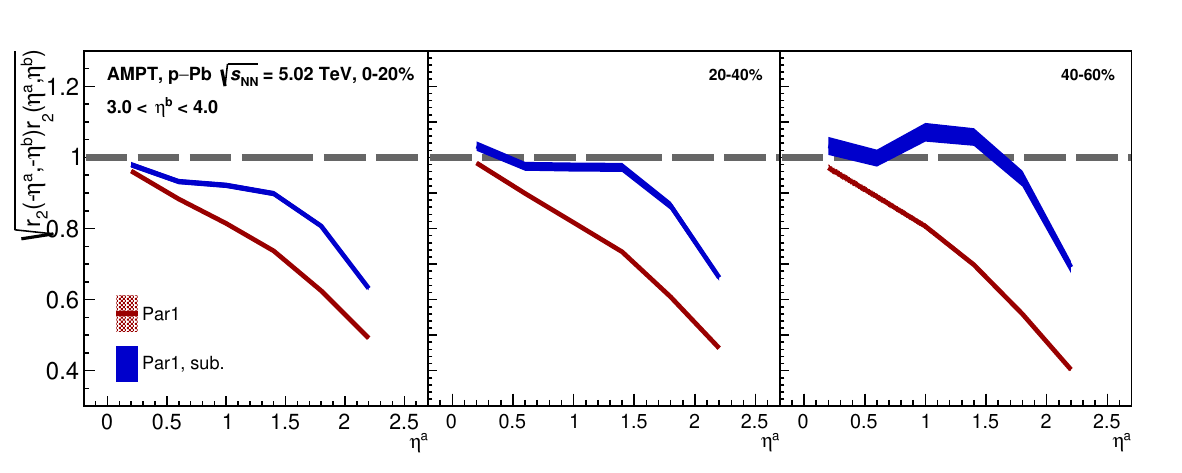}
\caption{(Color online) The comparison of the $\sqrt{r_{2}(\eta^{a},\eta^{b})r_{2}(-\eta^{a},-\eta^{b})}$, as a function of $\eta^{a}$ for 3.0 $< \eta^{b} <$ 4.0, in p--Pb collisions at 5.02 TeV between before and after nonflow subtraction. The results in 0--20\% (left), 20--40\% (middle) and 40--60\% (right) centrality classes are all shown.}
\label{Fig: eta decorr sub}
\end{center}
\end{figure*}

As defined in Eq.~\ref{eq: rn_eta_approx_pPb}, the $\eta$-dependent $\sqrt{r_{2}(\eta^{a},\eta^{b})r_{2}(-\eta^{a},-\eta^{b})}$ in p--Pb collisions at 5.02 TeV for 3 $< \eta^{b} <$ 4 and $4.4 < \eta^{b} < 5.0$ are shown in Fig.~\ref{Fig: eta decorr}. The results obtained from the AMPT with all configurations are compared to the measurements from the CMS collaboration. All calculations show a larger breakdown of factorization as $\eta^{a}$ increases, consistent with the data. Similar to the findings in the $p_{\mathrm{T}}$-dependent flow vector decorrelation, a significant breakdown of factorization is observed when the parton scattering cross section is set to 0 (Par3), but the deviation between the results of cross section $\sigma=0.5$ mb (Par1) and $\sigma=3$ mb (Par2) is small. Unlike the integrated-$v_{2}$ in Fig.~\ref{Fig: v2 eta}, the decorrelations obtained without hadronic scatterings (Par6) are consistent with Par1 calculations, suggesting that $\eta$-dependent decorrelation has no sensitivity to the hadronic scatterings. No significant effect of the resonance decay is observed for $\eta^{a} <$ 1, while a slight enhancement is shown for $\eta^{a} >$ 1 (Par1 vs Par5). On the other hand, varying the Lund parameters to $a=0.5$, $b=0.9$ (Par4) leads to a smaller decorrelation for both 3 $< \eta^{b} <$ 4 and 4.4 $< \eta^{b} <$ 5 (Par4 vs Par1), reflecting the effects from initial conditions. In general, the model calculations with all tunings, except for $\sigma=0$, provide a fair description of the data for $4.4 < \eta^{b} < 5.0$, but systematically overestimate the deviation from unity of decorrelations for 3 $< \eta^{b} <$ 4 and $\eta^{a}>$ 1.

Figure~\ref{Fig: eta decorr sub} shows the $\sqrt{r_{2}(\eta^{a},\eta^{b})r_{2}(-\eta^{a},-\eta^{b})}$ as a function of $\eta^{a}$ in 0--20\%, 20--40\% and 40--60\% centrality classes before and after the nonflow subtraction. The template fit method described in Eq.~\ref{eq: template fit} is applied for the calculations of Eq.~\ref{eq: rn_eta} and Eq.~\ref{eq: rn_eta_approx_pPb}, where the contribution from the long-range jet correlations is suppressed. A significant enhancement for the decorrelation is observed for the results after the subtraction in three centrality classes, indicating that the $\eta$-dependent flow vector decorrelation has a strong dependence on the nonflow (e.g., the dijet) in p--Pb collisions. This is consistent with the findings in recent measurements in even smaller pp collisions performed by the ATLAS collaboration~\cite{ATLAS:2023rbh}. On the other hand, the deviation between the results before and after the nonflow subtraction becomes larger with the increasing centrality. Especially in the 40--60\% centrality class, the tendency of the decorrelation completely changes after the nonflow subtraction, first increasing then decreasing with the increasing $\eta^{a}$. It is worth noting that such an effect from the long-range jet correlations is negligible for the $p_{\mathrm{T}}$-dependent decorrelation at mid-rapidity ($|\eta|<$0.8), as shown in Fig.~\ref{Fig: v2 sub comp}. This is because the contribution from long-range jet correlations to the $v_2$ is more significant at forward/backward rapidity compared to mid-rapidity in small collision systems~\cite{ALICE:2022ruh}. This hints at the important role of the long-range jet correlations in the longitudinal fluctuations and decorrelations of the flow vector.    
       
\section{SUMMARY}
\label{summary}
The decorrelation of the flow vector as well as its $\mathrm{p}_{T}$ and $\eta$ dependence in p--Pb collisions at $\sqrt{s_{\mathrm{NN}}}=$ 5.02 TeV are systematically studied with the AMPT model in this paper. The sensitivity of the decorrelations to the initial conditions, the partonic scatterings, the haronic rescatterings are probed by varying the configurations of the model. It is found that the observables are driven by the partonic scatterings and the initial conditions together, but has weak dependence on the hadronic rescattering. It suggests that the event-by-event fluctuations of the initial state lead to the fluctuations of the flow vector, then they are propagated to the observed decorrelation in final state according to the parton interactions. These results point to the possibility for employing flow vector decorrelation to investigate the transport properties of the hot and dense medium likely formed in small collision systems. In addition, we demonstrate that the contribution from the long-range jet correlations plays an important role in generating the longitudinal decorrelation in p--Pb collisions. It provides further insights in understanding the fluctuations of the flow vector in small collision systems and has referential value for future measurements.

\section{Acknowledgement}
This work was supported by National Natural Science Foundation of China (No. 12147101, 12405158, 12405159), the Natural Science Foundation of Hubei Province (No. 2024AFB136), the Fundamental Research Funds for the Central Universities, China University of Geosciences (Wuhan) with No.G1323523064, Key Laboratory of Quark and Lepton Physics in Central China Normal University (No. QLPL202106, QLPL2022P01, QLPL2023P01, QLPL2024P01). 

\bibliography{reference}

\end{document}